\documentstyle{elsart}
\newcommand{\slaninaefi}[1]{#1}
\begin{document}
\begin{frontmatter}
\title{Weak pinning: Surface growth in presence of a defect}
\author[Fzu,CTS]{F.~Slanina \thanksref{e-Franta}}
\thanks[e-Franta]{e-mail: slanina@fzu.cz}
\author[Fzu]{and M.~Kotrla  \thanksref{e-Mirek}}
\thanks[e-Mirek]{e-mail: kotrla@fzu.cz}
\address[Fzu]{
 Institute of Physics, 
 Academy of Sciences of the Czech Republic,\\
 Na~Slovance~2, 
 CZ-18040~Praha,
 Czech Republic
}
\address[CTS]{
 Center for Theoretical Study, 
 Jilsk\'a~1,
 CZ-11000~Praha, 
 Czech Republic 
}
\begin{abstract}
%
%
We study the influence of a point defect on the profile
of a growing surface
in the single-step growth model.
We employ the mapping to the asymmetric exclusion model with blockage,
and using Bethe-Ansatz eigenfunctions as a starting approximation 
we are able to solve this problem analytically in two-particle sector.
The dip caused by the defect is computed.
A simple renormalization group-like 
argument enables to study scaling of the dip with increasing
length of the sample $L$; the RG mapping is calculated approximately
using the analytical results for  
small samples.
For a horizontal surface we found
that the surface is
only weakly pinned at the inhomogeneity; 
the dip scales as a power law $L^\gamma$ with $\gamma= 0.58496$.
The value of the exponent agrees with direct
numerical simulations 
of the inhomogeneous single-step growth model.
In the case of tilted surfaces we observe a phase transition
between weak and strong pinning and the exponent in the weak pinning
regime depends on the tilt.
\end{abstract}


\end{frontmatter}
%
%
%
%
\section{Introduction}
There has been much activity in the field of surface growth in the
last period
\cite{list_krugetall,list_levietall}.
Some problems are now rather well understood,
nevertheless, many questions still remain open or  not 
completely clarified. 
One of them is the effect of defects
on growth. In particular,
we have in mind a defect 
which 
persists during the growth and makes the growth inhomogeneous.
This problem is relevant for various physical situations.
For example, 
1+1 dimensional epitaxial growth (one-dimensional substrate and
one dimension in the direction of growth)
of steps on vicinal surfaces when there is a line defect on the terrace
in the growth direction.
Growth becomes inhomogeneous,
the growth velocity at the defect can be lower  
(the moving step edge is pinned to the defect)
or higher (the defect causes an excess of incorporated particles)
than averaged velocity of the homogeneous step edge.
An impurity floating on the growth front 
can have similar effect.
In the case of 2+1 dimensional growth
on a substrate with a defect (for example a dislocation) 
the defect is replicated in the grown material
and makes the growth inhomogeneous. 
Still another situation
is growth of a two-component material forming two domains separated
by a domain wall; growth rate at the domain wall is different from
growth rate within domains \cite{ko_pre_97}.
Epitaxial growth in these situations was so far little studied theoretically,
although it is of great practical importance.
Nevertheless, similar problems were investigated in
different contexts 
\cite{wo_ta_90,ka_mu_92,ta_lyu_93,de_ev_pa_93,ka_ge_mu_de_93,to_zia_96}. 

From the statistical-mechanical point of view
the central problem is the existence and the character of 
pinning - depinning transition. The lack of deposition
may result in a dip of depth $d$ located at the defect. The growing
surface may be pinned, if $d$ scales linearly with the sample size $L$, or it
may be depinned, if $d$ remains finite when the sample size goes to
infinity. However, the dip may scale as a power of $L$, $d\sim
L^\gamma$. 
To distinguish between different situations
we shall call the case with $\gamma=1$ {\it strong pinning}, and 
the case $0<\gamma<1$ {\it weak pinning}. The situation when $\gamma=0$
corresponds either to depinning, if 
$\lim_{L\to\infty}d={\rm const}<\infty$, or to
various types of logarithmic pinning, like $d\sim (\log L)^\nu$ with
positive exponent $\nu$ .

Several authors investigated the problem  of pinning of growing
interfaces 
through various formulations.
Wolf and Tang \cite{wo_ta_90} studied inhomogeneous growth analytically 
using the growth equation proposed by Kardar, Parisi and 
Zhang (KPZ) \cite{ka_pa_zha_86}. 
They found that depending on the sign of the coefficient $\lambda$ in front
of the non-linear term in the KPZ equation
there is different behaviour of the surface profile with the size $L$ of the
sample for the lack and for the excess of deposition at the defect.
For $\lambda >0$, extra deposition at the defect site
leads to a pile of amplitude $d$ increasing linearly with
$L$, whereas the lack of deposition produce a groove of depth $d$,
increasing logarithmically, $d\sim \log L$.
For $\lambda<0$, linear and logarithmic dependence are interchanged.

Kandel and Mukamel \cite{ka_mu_92} performed 
numerical simulations of 1+1 dimensional 
polynuclear growth model  with a defect.
They observed a defect-induced phase transition when changing the
strength of the defect, manifested by the change of the dependence
of the dip, $d$, on the system size. 
They found that $d$ scales as
$d\sim L^\gamma$, with $\gamma=1$ in the strong-defect phase, while
$\gamma < 1$ for weak defect phase.  
In further investigation, a mean-field description of the phase transition was
used \cite{ka_ge_mu_de_93} and the mean-field value of the
exponent $\gamma$ 
in the weak defect phase was found to be $\gamma=0$.

Tang and Lyuksyutov \cite{ta_lyu_93} investigated
the problem in mathematically equivalent formulation as
localization of a directed polymer in disordered medium.
They found that 
there is no depinning transition in 1+1 dimensions for a line
defect\footnote{In the context of directed polymers the term line
defect corresponds to what we shall call point defect in the growth
model or asymmetric exclusion model terminology.}  
and
in 2+1 dimensions for a plane defect, while depinning occurs for a line defect
in 2+1 dimensions. This finding is not in contradiction with the results
of \cite{ka_mu_92}, because in \cite{ta_lyu_93} the short-time regime $L\gg
t^\zeta$ is observed, where the weak pinning cannot be distinguished from 
strong pinning. 

The problem of growth in the single-step model
is equivalent to diffusion in  the asymmetric exclusion model 
\cite{de_ev_pa_93},
where the defect corresponds to a blockage.
Such a situation was already studied analytically and some results on
the nature of the impurity-induced phase transition are
available. They deal with the presence of a strongly pinned phase.
The matrix technique developed by Derrida {\it et al. } \cite{de_ev_pa_93} for
studying the asymmetric exclusion model with open boundary conditions
was used in the case of mobile impurity in the medium, observing a
phase transition when changing the strength of the impurity 
\cite{mallick_96}. 

Using Bethe Ansatz it was also argued in favour of the existence of the phase
transition induced by a point defect \cite{he_schu_94}. 
The pinning-depinning transition corresponds to the phase transition
characterized by condensation of particles.
The exact analysis of small systems ($L\le 4$) accompanied by the use
of Pad\'e
approximants \cite{ja_le_94} gives some rigorous bounds on the region
of strongly pinned phase. However, the possibility of the phase
transition to the weak pinning regime remained open.

To our knowledge, the case of weak pinning was not explicitly
considered in the existing literature and  none of available 
calculations yields reasonable estimate of the 
exponent $\gamma$ in the weakly pinned phase, even though the matrix
technique should in principle be able to give exact solution and the
generalized Bethe Ansatz for complicated boundary conditions seems
promising \cite{schutz_93}. 
At present it is not clear under which conditions the weakly pinned
phase with  
$0<\gamma<1$ exists and what is the value of $\gamma$.

In this paper we 
show the existence of weak pinning and present the calculations of
the exponent $\gamma$ in 1+1 dimensional single-step model of
inhomogeneous growth. We use an approach combining Bethe Ansatz 
and renormal\-ization-group
argument. 
We also present results of numerical simulations
confirming the predicted exponent.

The article is organized as follows.
In section 2 we describe the growth model considered, and give the
definition of the dip for the surface of an arbitrary
orientation. In 
section 3 the one particle solution is described explicitly. The latter
result is used in section 4 for construction of two-particle states.
Section 5 contains our main results, both analytical and numerical,
and the 
last, sixth section is devoted to discussion of the results.

\section{Growth model with a local defect}
In  numerical simulations of surface growth,
solid-on-solid (SOS) models are often employed because the surface
can be simply described by a single height function $h(x)$
of the substrate coordinate $x$.
In most of numerical studies of inhomogeneous growth,
so-called restricted solid-on-solid (RSOS) models with the additional
constraint on the height difference of the neighbouring sites
were used. 
The direct consequence of the constraint is that 
the number of possible configurations is finite.
The RSOS models are rather well understood in the case of 
homogeneous growth \cite{list_krugetall}. 
The simplest one of these models is the single-step model
with the height difference being only $+1$ or $-1$.
This class of models is 
particularly suitable for numerical as well as analytical approaches.
In 1+1 dimensions an exact solution is known, 
using the Bethe Ansatz 
\cite{list_dharetall}. 

We consider an inhomogeneous 1+1 dimensional single-step growth model.
The state of the system is described by an array
$h(x),x=$ 0,1,2,...,$L$ 
of integers, which represent the height of the
surface at the site $x$ with the constraint
$|h(x+1)-h(x)|= 1$ 
which enables representation of the state of the
surface using Ising spin variables $\sigma(x)$
defined by 
$\sigma(x)= h(x)-h(x-1)$. 
We impose cyclic boundary conditions. In terms of the spin variables
they read 
$\sigma(0)= \sigma(L)$, while for the height variables they fix the
difference $h(0)-h(L)$, {\it i. e. } the global tilt, constant during
the growth. 
We consider only pure growth (no evaporation).
Growth is possible only at a growth site $x$ which obeys
the condition
$\sigma(x)=-1, \sigma(x+1)=1$. The deposition of a particle results in
the change of the configuration:
$h(x)\to h(x)+ 2$ {\it i. e. } 
$\sigma(x)\to -\sigma(x), \sigma(x+1)\to -\sigma(x+1)$.
The inhomogeneity is introduced 
by the change of the rate for deposition on one fixed site.
We suppose the 
deposition rate at a growth site $x\ne x_0$ to be equal to 1, and the 
deposition rate at a growth site on the defect site $x=x_0$ to be $1-\alpha$ .
The variable $\alpha$ denotes the strength of the defect.
The value $\alpha=1$ corresponds to total 
inhibition of growth at the defect, {\it i.e. } to maximum possible
pinning; 
due to the single-step constraint the surface profile evolves
to a triangular
shape in which it remains frozen.

Let $h(x,t)$ be the
surface height at time $t$. To measure the pinning of the surface
with an arbitrary tilt $(h(L,t)-h(0,t))/L$, we introduce the quantity
\begin{equation}
 \Delta\bar{h}(x)={1 \over T}\sum_{t=t_0}^{t_0+T}
 \left(
  h(x,t)-h(0,t)-
  x{h(L,t)-h(0,t)\over L}
 \right)\; ,
\label{height}
\end{equation}
which we call averaged height difference. It is
the height averaged over (long) time interval $T$ with the steady
increment of the height and the tilt of the surface 
subtracted. If there were no defect, 
the averaged height difference would be zero. 
The defect at $x_0$  introduces an 
inhomogeneity into the system and $\Delta\bar{h}(x)$ may 
depend on
position. We define the dip at the defect site as
\begin{equation}
\label{eq:dip}
d={1\over L}\sum_{x=1}^L \Delta \bar{h}(x)-\Delta\bar{h}(x_0)\; .
\end{equation}
The surface is pinned, if $d/L$ remains finite in the
thermodynamic limit, $L\to\infty$. It seems natural to say that the
surface is depinned if $d/L\to 0$. In this sense,
$\lim_{L\to\infty}\frac{d}{L}$ is the order 
parameter of the pinning transition. However, $d/L$ may converge to
zero in the thermodynamic limit
even if $d\sim L^\gamma$ with $0<\gamma<1$. The dip blows up, but more
slowly than the length of the 
sample, therefore we call this situation {\it weak pinning } with exponent
$\gamma$.

The dynamics of the system has the following conservation law. The sum
of all spin variables, $S = \sum_x \sigma(x)$ does not change during the
growth. $S$ should be understood as  the average orientation of the
surface, $S=h(L)-h(0)$, fixed by the boundary conditions. The
single-step model is mapped to the asymmetric 
exclusion model if we 
consider down spins as particles and up spins as unoccupied sites. In
the language of asymmetric exclusion model, the corresponding
conserved quantity is the 
number of particles $n=(L-S)/2$. The particle density $\rho = n/L$ is
$1/2$ for horizontal and $\rho\ne 1/2$ for tilted surface.

The situation corresponds to
directed diffusion of $n$ 
hard core particles on a line. At each time step, each particle can
remain in the original 
position or jump to its right next neighbour site, provided it is
unoccupied. The jumping rate from the site 
$x_0$ to $x_0+1$ is $1-\alpha$, while the jumping rate from $x$ to
$x+1$ for $x\ne x_0$ is 1.

The conservation law facilitates the exact solution of the system. The Bethe
Ansatz solution
\cite{list_dharetall}
 starts  
with writing the master equation for the probabilities of all possible
configurations. The set of  all configurations can be divided into sectors
according to number of particles. For each number of particles $n$,
the master equation
\begin{equation}
\dot{p}(x_1,x_2,...,x_n)= \sum_{\{x^\prime\}} T_{n\{x\}\{x^\prime\}}
p(x_1^\prime,...,x_n^\prime)
\end{equation}
is solved independently. Here, the variables $x_1,...,x_n$ describe
positions of the 
particles and $p(x_1,...,x_n)$ is probability of finding the particles
in the specified spatial configuration. The spectrum of the
$n$-particle transition 
matrix $T_n$ contains all information on the dynamics of the system.
The solution for arbitrary $n$ is still not known. We shall present
exact solution for $n=1,2$.

\section{One-particle problem}
In the case of a single diffusing particle the solution is quite easy.
For $n=1$ the master equation has a very simple form
\begin{equation}\left.\begin{array}{llll}
\dot{p}(x)&=&p(x-1)-p(x)&\;\;,\;{\rm for }\;\; x\neq x_0,x\neq x_0+1\;\;,\\
\dot{p}(x_0)&=&p(x_0-1)-(1-\alpha)p(x_0)&\;\; ,\\
\dot{p}(x_0+1)&=&(1-\alpha)p(x_0)-p(x_0+1)& \;\; .
\end{array}\right.\end{equation}
We can solve it by finding the (right) eigenvectors of the transfer
matrix $T_1$ in the one-particle sector. The eigenvectors are parametrized
by 
a complex number $\zeta$. As we shall see, also the left
eigenvectors are useful, so we determine both of them.

We write
\begin{equation}
\sum_y T_{1xy}\pi_\zeta(y)=\lambda_\zeta \pi_\zeta(x)
\end{equation}
for the right eigenvectors and
\begin{equation}
\sum_x \pi_\zeta^T(x) T_{1xy}=\lambda_\zeta\pi_\zeta^T(y)
\end{equation}
for the left ones. The left and right eigenvectors for different
$\lambda_\zeta$ are mutually orthogonal.

We look for the solution in the form
$\pi_\zeta(x)=A\zeta^x$
for $x<x_0$ or $x>x_0$,
where $A$ may be different in the regions $x<x_0$ and $x>x_0$.
We find easily
\begin{equation}\left.\begin{array}{lll}
\label{eq:pi}
\pi_\zeta(x)=    
&\zeta^x                         
&\; ,\;{\rm for }\; x<x_0\;\;,\\
\pi_\zeta(x_0)=  
&\zeta^{x_0}{1\over 1-\alpha \zeta}  
&\; ,\\
\pi_\zeta(x)=    
&\zeta^x{1-\alpha\over 1-\alpha \zeta} 
&\; ,\;{\rm for } \; x>x_0\;\;,
\end{array}\right.\end{equation}
and similarly for the left eigenvectors
\begin{equation}\left.\begin{array}{lll}
\label{eq:piT}
\pi^T_\zeta(x)=    
&\zeta^{-x}                           
&\;,\;{\rm for }\;x\le x_0\;\; ,\\
\pi^T_\zeta(x)=    
&\zeta^{-x}{1-\alpha \zeta\over 1-\alpha} 
&\;,\;{\rm for }\;x>x_0\;\; .
\end{array}\right.\end{equation}
The corresponding eigenvalue is $\lambda_\zeta=\zeta^{-1}-1$.
The stationary solution $p_s(x)$ of the master equation is the
eigenvector with zero eigenvalue, {\it i.e. } with $\zeta=1$, normalized so
that 
$\sum_x p_s(x)=1$. 
The stationary solution gives immediately the stationary height profile.
For $\alpha=0$, the value of $\Delta\bar{h}(x)$ is zero
for all $x$,
whereas for $\alpha>0$ we have
\begin{equation}\left.\begin{array}{lll}
 \Delta\bar{h}(x)=
 &{2\alpha x\over L(L(1-\alpha)+\alpha)}
 &\; ,\;{\rm for}\;\;x<x_0\;\; ,\\
 \Delta\bar{h}(x)=
 &{2\alpha (x-L)\over L(L(1-\alpha)+\alpha)}
 &\; ,\;{\rm for }\;\;x\ge x_0\;\; .
\end{array}\right.\end{equation}
Hence the dip at the defect is
\begin{equation}
 \label{dip_1}
 d(1,L,\alpha)=
 \alpha {L-1\over L(L(1-\alpha)+\alpha)}\;\; .
\end{equation}

Because there is only one particle, it is obvious that the surface is
not pinned; the value of the dip for 
all $\alpha$ and $L$ is
at most 1. So, in
order to actually observe the pinning, we should have non-zero density
of particles in the thermodynamic limit $L\to\infty,n\to\infty$. This
requires a 
substantially different approach.

\section{Lattice constant doubling}
We apply a scheme similar to the real space renormalization group
technique. Suppose 
we have even number $n$ of particles in the sample of length
$L$, which is also even. The time evolution is governed by the
transfer matrix 
$T_n(\alpha)$. (From now on we shall write explicitly the dependence
of the 
transfer matrix on $\alpha$). The presence of the defect of strength
$\alpha$ results in a dip 
$d(n,L,\alpha)$. Taking each pair of neighbour particles as a single
object, we have system of $n/2$ effective particles. Their
interactions are 
rather complex, but the fact, that the particles are hard-core, {\it i.e. }
they can not skip one another, is preserved even for the effective
particles. The time evolution is no more described by a Markov process
governed by some transfer matrix, but by a more general evolution
operator ${\cal T}_{n/2}(\alpha)$. However, we suppose that we can
approximate the dynamics by a Markov process and map the system of
$n/2$ effective particles onto the original system, but with half number
of particles and different effective strength of the defect $\alpha_2$:
\begin{equation}
 {\cal T}_{n/2}(\alpha)\to T_{n/2}(\alpha_2)\;\; .
\end{equation}

The lattice constant of the new system is twice as large, because
pairs of 
particles occupy two original sites, so the length of the rescaled
sample is half of the original one. Also the height is rescaled by a 
factor 2. The procedure roughly corresponds to replacing four atoms
deposited 
independently on the surface by a single ``composite'' atom made of
four original atoms glued together.  The dependence of rescaled
strength of the defect 
$\alpha_2$ on the original $\alpha$ can be computed by equating the 
dip in the original and rescaled system
\begin{equation}
 \label{compare_dips}
 d(n,L,\alpha)=
 2d(n/2,L/2,\alpha_2)\;\; .
\end{equation}
This equation may be understood also as the requirement of the
equality of deposited mass in the original system and the rescaled
system. 

A justification for the doubling procedure may be provided {\it a
posteriori} by the comparison of 
the whole surface profiles in the original and rescaled systems. If the
deviation is small even far from the location of the defect, we
can suppose that the pairs of particles actually behave like the
original particles, but in rescaled system.

The simplest way for obtaining an approximation for the function
$\alpha_2(\alpha)$ is to 
compute the average height profile for $n=2$ and the corresponding
dip. From (\ref{dip_1}) and (\ref{compare_dips}) it follows that
\begin{equation}
 \alpha_2=
 \left({L^2\over L-2}\right) 
 {d(2,L,\alpha)\over 4+Ld(2,L,\alpha)}\;\; . 
\end{equation}

Fortunately, the $n=2$ case is soluble exactly.
If $\alpha=0$, the solution is given simply by Bethe Ansatz. The
many-particle eigenvector is the linear combination of products of
one-particle 
eigenvectors. However, doing the same for $\alpha\ne 0$ does not work,
because the system lacks translational symmetry. But the 
Bethe Ansatz eigenfunctions $p_B$ may serve as a good starting point for
further calculations, because they must obey the equation
\begin{equation}
 \label{secular_2}
 T_2(\alpha)p_B =
 \lambda p_B
\end{equation}
in all cases when both particles are far from the defect.
Moreover, if we construct the eigenfunction from the exact one-particle
eigenvectors (\ref{eq:pi}), which are calculated for given
$\alpha\ne0$, 
then the 
equation (\ref{secular_2}) will hold even if one of the particles is
at the defect. Only such terms when both particles are at (or
near) the defect will disagree. But we may treat the difference as a
localized perturbation and solve exactly the equation for the
resolvent $G(z)=(z-T_2)^{-1}$.

So, let us start by explicitly writing the eigenvalue equation
(\ref{secular_2}). We define the hopping rate from the site $x$
\begin{equation}
w_x=1-\alpha\,\delta_{xx_0}\;\; .
\end{equation}
Then, the two-particle eigenvalue equation is
\begin{equation}\left.\begin{array}{lll}
 \label{secular_2off}
 \lambda\; p(x,y)&=   &w_{x-1}\,p(x-1,y)+w_{y-1}\,p(x,y-1)\\
 &&-(w_x+w_y)\,p(x,y)\;\;,\\
 \lambda\; p(x,x+1)&= &w_{x-1}\,p(x-1,x+1)-w_{x+1}\,p(x,x+1)\;\; .
\end{array}\right.\end{equation}
From the cyclic boundary conditions it follows that $p(x,y+L)=p(y,x)$.

We construct the Bethe eigenfunctions
\begin{equation}
p_{B\zeta\omega}(x,y)=
\pi_\zeta(x)\pi_\omega(y)+A_B\pi_\omega(x)\pi_\zeta(y)\;\; ,
\end{equation}
using the one-particle functions (\ref{eq:pi}). The constant
\begin{equation}
A_B=-{\omega-1\over \zeta-1}
\end{equation}
is set so that (\ref{secular_2off}) are satisfied far from the defect
\cite{list_dharetall}. 
For $\zeta=\omega=1$ we take $A_B=0$. The corresponding 
$\lambda$ is 
\begin{equation}
\lambda=\lambda_{B\zeta\omega}=\zeta^{-1}+\omega^{-1}-2\;\; .
\end{equation}
The values of $\zeta$ and $\omega$ are set by the cyclic boundary
conditions, resulting in a pair of algebraic equations
\begin{equation}\left.\begin{array}{lll}
 \zeta^L=-{\zeta-1\over \omega-1}{1-\alpha \zeta\over 1-\alpha}&\;\; ,\\
 \omega^L=-{\omega-1\over \zeta-1}{1-\alpha \omega\over 1-\alpha}&\;\; ,
\end{array}\right.\end{equation}
which can be solved numerically by standard methods. 
Together with the solution 
$\zeta=\omega=1$ 
we have $L(L-1)/2$ pairs $(\zeta,\omega)$. 
Each of them determines one 
of right eigenvectors of some matrix $T_{2B}$, which differs from
$T_2$ only in several matrix elements, so we can take $T_{2B}$ as a
first approximation to $T_2$ and treat the difference $T_2-T_{2B}$ as
a perturbation. Let $P$ be matrix whose columns are
the Bethe eigenfunctions $p_{B\zeta\omega}(x,y)$ (pairs $(x,y)$ denote rows,
pairs $(\zeta,\omega)$ denote columns). We can write
\begin{equation}
 T_{2B}=P^{-1}\Lambda P
\end{equation}
where $\Lambda$ is the diagonal matrix with numbers
$\lambda_{B\zeta\omega}$ on 
the diagonal. It is quite easy to find the inverse $P^{-1}$ if
$\alpha=0$. As Bethe functions $p_B$ are exact right eigenvectors of
$T_2$, the Bethe functions $p_{B\zeta\omega}^T$ constructed by the
same 
procedure from the left one-particle eigenvectors $\pi_\zeta^T$ are
exact 
left eigenvectors of $T_2$. Because left and right eigenvectors with
different eigenvalues are orthogonal, the matrix composed of properly
normalized left
eigenvectors is the inverse of $P$. When $\alpha\ne 0$, the
situation is different, but in fact we need to make only
a tiny modification.

It can be checked directly (actually we performed the check using
Maple V) that the vectors
\begin{equation}\left.\begin{array}{lll}
 p^T_B(x,y)&=
 &\pi^T_\omega(x)\pi^T_\zeta(y)+A_B\pi^T_\zeta(x)\pi^T_\omega(y)\;\;,\\
 &&\;\;\;\;\;\;\;\;\;\;\;\;\;\;{\rm for }\;\; (x,y)\ne(x_0,x_0+1)\;\;,\\
 p^T_B(x_0,x_0+1)&=
 &{1\over 1-\alpha}\left(\omega^{-x_0}\zeta^{-x_0-1}
 +A_B\zeta^{-x_0}\omega^{-x_0-1}\right)
\end{array}\right.\end{equation}
are orthogonal to the vectors $p_B$ and so, after proper
normalization, they represent rows of the matrix 
$P^{-1}$. 

As we already said, vectors $p_B$ are considered as nearly exact
eigenvectors of $T_2$. What we exactly mean is that
\begin{equation}
T_2 P=P\Lambda+\Delta\;\; ,
\end{equation}
where the matrix $\Delta$ has only two non-zero rows, corresponding to
row indices $a=(x_0-1,x_0)$ and $b=(x_0,x_0+1)$. (The column indices are
the pairs $(\zeta,\omega)$).
We have explicitly
\begin{equation}\left.\begin{array}{lll}
\Delta_{a(\zeta,\omega)}&=&
\alpha(\zeta^{-1}+\omega^{-1}-1-\alpha) {\zeta-\omega\over\zeta-1} 
{(\zeta\omega)^{x_0}\over (1-\alpha\zeta)(1-\alpha\omega)}\;\;,\\
\Delta_{b(\zeta,\omega)}&=&
-\alpha(1-\alpha){\zeta-\omega\over\zeta-1} 
{(\zeta\omega)^{x_0}\over (1-\alpha\zeta)(1-\alpha\omega)}\;\;,\\
\Delta_{a(1,1)}&=&
{\alpha\over 1-\alpha}\;\;,\\
\Delta_{b(1,1)}&=&
-{\alpha\over 1-\alpha}\;\; .
\end{array}\right.\end{equation}

The solution is obtained, as soon as we compute the resolvent \linebreak
$G(z)=$ $(z-T_2)^{-1}$. If we define the ``unperturbed'' resolvent
$G^o(z)=$ $(z-P^{-1}\Lambda P)^{-1}$, 
using elementary matrix algebra we obtain equation
\begin{equation}
G(z)=G^o(z)+G(z)\Delta (z-\Lambda)^{-1}P^{-1}\;\; ,
\end{equation} 
which we call Dyson equation, taking an analogy from the quantum
mechanics. 

Let $\Pi$ be projector to the two-dimensional subspace corresponding
to indices $a$ 
and $b$ and $\bar{\Pi}=1-\Pi$ the complementary projector to it. 
The Dyson equation can be solved within that subspace,
because $\Pi\Delta=\Delta$. We obtain
\begin{equation}\left.\begin{array}{lll}
G(z)\Pi & =
&G^o(z)\Pi(1-\Delta(z-\Lambda)^{-1}P^{-1}\Pi)^{-1}\\
G(z)\bar{\Pi} & =
&G^o(z)\bar{\Pi}
+G^o(z)\Pi(1-\Delta(z-\Lambda)^{-1}P^{-1}\Pi)^{-1}\times\\
&&\times\Delta(z-\Lambda)^{-1}P^{-1}\bar{\Pi}\;\; .
\end{array}\right.\end{equation}
Computation of the matrix inverses is very simple, because they
involve only matrices $2\times 2$.

The stationary state is obtained from the resolvent in the $z\to 0$
limit
\begin{equation}
p_s(x,y)=\lim_{z\to 0} zG_{(x,y)a}(z)\;\; .
\end{equation}
The column index of the resolvent may be chosen arbitrarily. It
corresponds to the fact that stationary state does not depend on
initial conditions.

\section{Results}

We calculated explicitly the two-particle ($n=2$) stationary state
$p_s(x,y)$, defined on 
the triangle $1\le x<y\le L$ for
$L$ ranging from 4 to 32. The figures 
\ref{fig:1} 
\slaninaefi{
\begin{figure}[hb]
  \centering
  \vspace*{80mm}
  \includegraphics{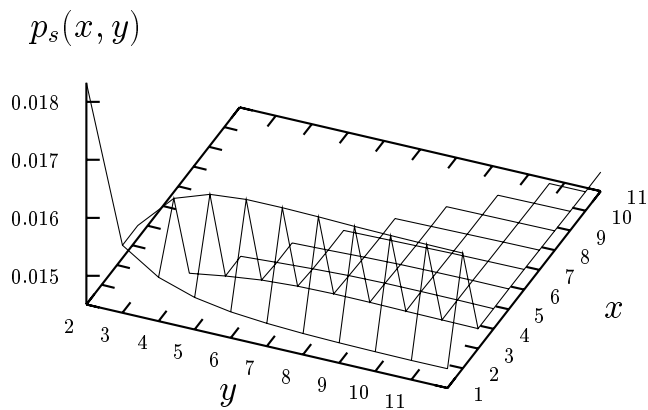}
  \caption{
        Stationary state of the two-particle 
        system for $L=12$ and $\alpha=0.1$.}
  \label{fig:1}
\end{figure}
}
and \ref{fig:2}
\slaninaefi{
\begin{figure}[hb]
  \centering
  \vspace*{80mm}
  \includegraphics{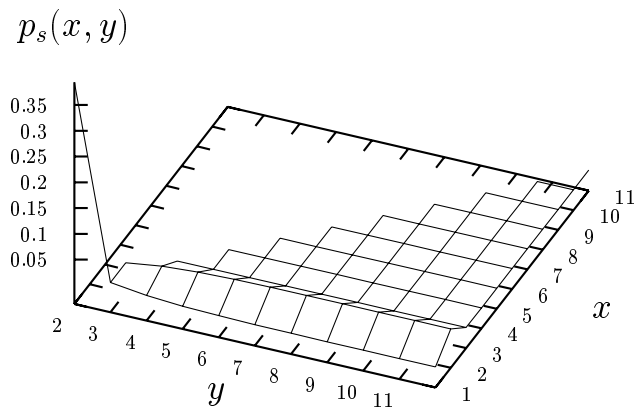}
  \caption{
        Stationary state of the two-particle 
        system for $L=12$ and $\alpha=0.9$.}
  \label{fig:2}
\end{figure}
}
show the stationary state for $L=12$ and $\alpha=0.1,0.9$. In both
cases, the defect is 
located at $x_0=2$. We can see that for small $\alpha$ the effect of the
defect is weaker, but more delocalized in the two-dimensional
configuration space $(x,y)$. The probability
is distributed nearly homogeneously, but there is a band of higher
probability composed of points which have any of the coordinates $x,y$
equal to $x_0$. 
Large $\alpha$ leads to localization of the eigenfunction around
the point $(x_0-1,x_0)$; in the latter case a single peak develops
which bears nearly all the probability.

These results were further used to 
calculate the stationary height profile difference $\Delta\bar{h}(x)$
and dependence $\alpha_2(\alpha)$ of the renormalized strength of the
defect on the bare value of $\alpha$. The figure 
\ref{fig:3} 
\slaninaefi{
\begin{figure}[hb]
   \centering
   \vspace*{100mm}
   \includegraphics{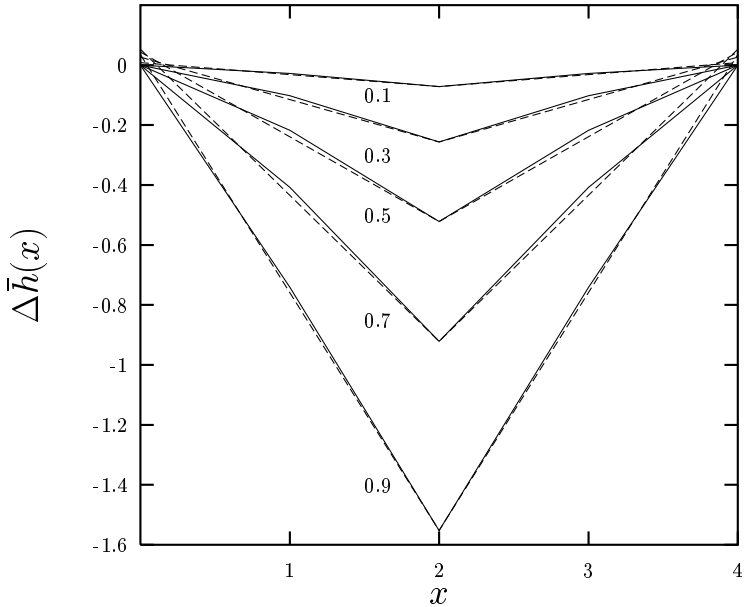}
   \caption{
         Stationary height difference for the two-particle
         system for $L=4$ and $\alpha = 0.1, 0.3, 0.5, 
         0.7, 0.9$ 
         (solid lines) together with effective 
         profile of corresponding one-particle system (dashed  
         lines). The numbers close to each full line denote 
         the corresponding $\alpha$.}
   \label{fig:3}
\end{figure}
}
shows the profiles for
$L=4$ and different $\alpha$'s. This situation corresponds to
horizontal surface ($\rho=1/2$). In
the same Fig. 
\ref{fig:3} 
we can also compare
the exact $n=2$ (two-particle) profiles with the effective
one-particle ones, calculated for 
$n=1$, $L=2$ and defect strength $\alpha_2(\alpha)$. We can see that
the 
agreement is rather good in the whole range of $\alpha$, which
justifies our doubling procedure. The figure
\ref{fig:4}  
\slaninaefi{
\begin{figure}[hb]
   \centering
   \vspace*{100mm}
   \includegraphics{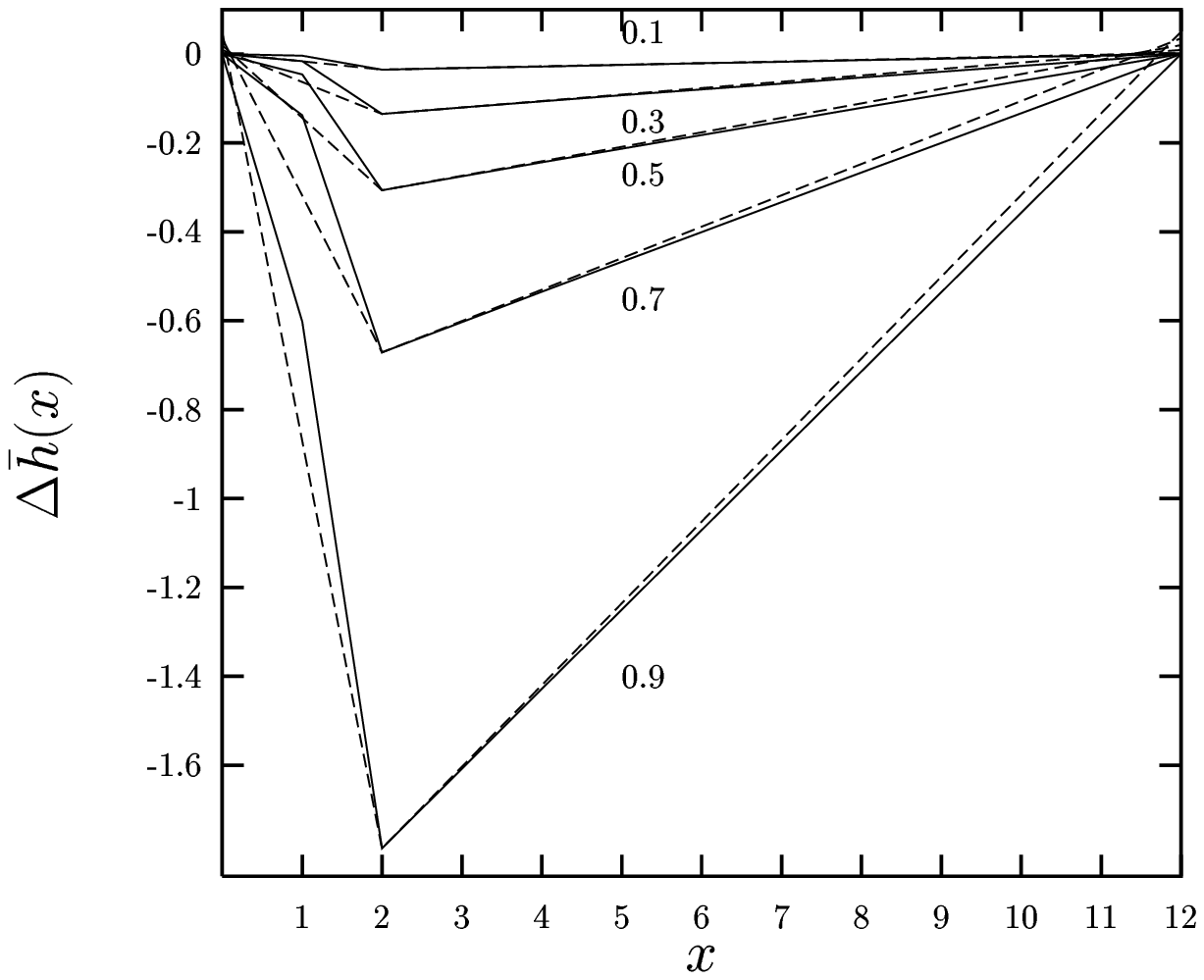}
   \caption{
         Stationary height difference, for the two-particle
         system for $L=12$ and $\alpha=0.1,0.3,0.5,0.7,0.9$  
         (solid lines) together with effective  
         profile of corresponding one-particle system (dashed 
         lines). The numbers close to each full line denote 
         the corresponding $\alpha$.}
   \label{fig:4}
\end{figure}
}
shows the same for $L=12$. These height 
profiles are tilted ($\rho<1/2$) which leads to the asymmetry of the
dip. Even in this case we can see good agreement
between the exact two-particle profile and the effective one-particle
one. However, for small $\alpha$ and near the defect, the agreement
is worse than for horizontal surface.

The key result of our computation is the function
$\alpha_2(\alpha)$. In the figure 
\ref{fig:5},
\slaninaefi{
\begin{figure}[hb]
   \centering
   \vspace*{100mm}
   \includegraphics{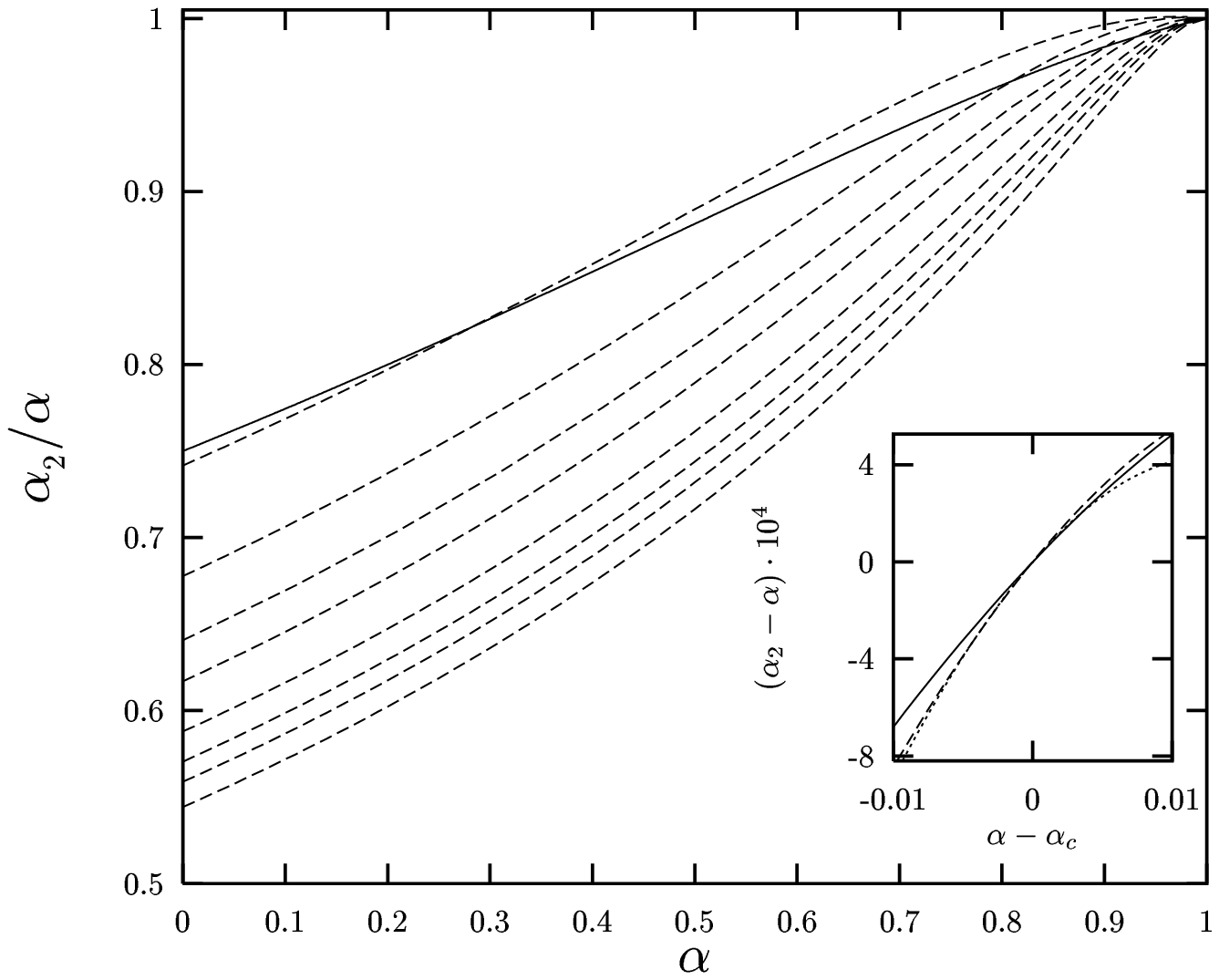}
   \caption{
         The $\alpha$-dependence of the ratio of 
         the effective defect strength $\alpha_{_2}$ to
         original $\alpha$ for $\rho=1/2$ (solid line) and
         $\rho=1/3,1/4,1/5,1/6
         ,1/8,1/10,1/12,1/16$
         (dashed lines, from top to bottom). In the inset, detail of
         the dependence of difference $\alpha_2-\alpha$ on the
         distance from the critical point, $\alpha-\alpha_c$, for
         $\rho=1/3$ (solid), $\rho=1/4$ (dashed), 
         $\rho=1/5$ (dotted).}
   \label{fig:5}
 \end{figure}
}
the dependence of $\alpha_2(\alpha)/\alpha$ on $\alpha$ is shown for
both horizontal ($\rho = 1/2$) and tilted ($\rho < 1/2$) surfaces.

The physics of pinning is contained in fixed points of the map 
$\alpha\to\alpha_2(\alpha)$. There are surely at least two fixed
points, namely 0 and 1, corresponding to weak pinning (or depinning)
and strong pinning regime, respectively. Their properties may be
determined from the quantities 
\begin{equation}\left.\begin{array}{lll}
s_0 & = & \lim_{\alpha\to 0}{\alpha_2\over\alpha}\;\; ,\\
s_1 & = & \lim_{\alpha\to 1}{1-\alpha_2\over 1-\alpha}\; .
\end{array}\right.\end{equation}

The inequality $s_i<1 (>1)$ means that the fixed point $\alpha = i$ is
stable (unstable).

Let us comment on the weak pinning regime first.
The value of
$s_0$ determines the scaling of the dip in the weak pinning
regime for $\alpha\to 0$. The exponent is
\begin{equation}
\gamma=
1+{\log s_0\over\log 2}\;\; ,
\end{equation} 
as can be seen from equations (\ref{dip_1}) and (\ref{compare_dips}).

The values we observe in the Fig. \ref{fig:5} are
$1>s_0>0.5$ in all cases, corresponding 
to weak pinning, $1>\gamma>0$, while $s_0\le 0.5$ would be signal of
depinning (or logarithmic pinning), $\gamma=0$.
Thus, the weak pinning regime is always present.

 The most
important result is the value of the exponent for horizontal profile.
We have observed an interesting result, that for $\rho=1/2$ it is $s_0=3/4$
up to 
the precision of our numerical procedure ($10^{-6}$). So, we
conjecture, that the 
exact value of the exponent for horizontal surface and $\alpha\to 0$ is 
\begin{equation}
\label{eq:gamma}
\gamma={\ln 3/2\over \ln 2}=0.58496...\;\; .
\end{equation}
This value may be compared directly with numerical simulations.

The dependence of $s_0$ and the corresponding exponent $\gamma$ on the
particle density $\rho=n/L$ is shown in Fig. 
\ref{fig:6}.
\slaninaefi{
\begin{figure}[hb]
   \centering
   \vspace*{80mm}
   \includegraphics{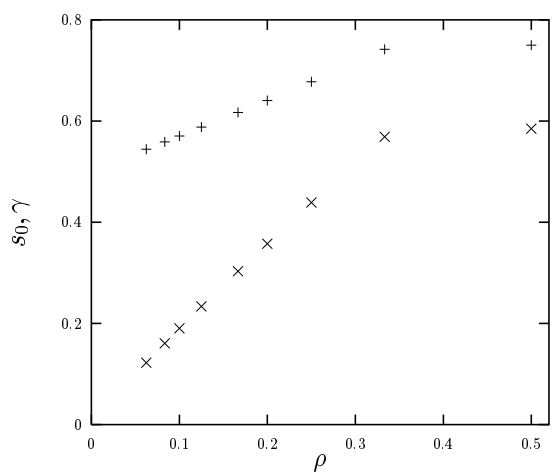}
   \caption{
         The slope $s_0$ at $\alpha\to 0$ ($+$)
         and the exponent $\gamma$ ($\times$) against the particle
         density $\rho=n/L$.}
   \label{fig:6}
\end{figure}
}
We can see that the exponent $\gamma$ decreases smoothly when
decreasing the density
and seems to approach 0 for $\rho\to 0$. It reflects the fact,
that for zero density there is only finite number of particles on
infinite sample and the dip should be also finite. The point at
$\rho=1/2$ does not fit to this smooth dependence. The explanation lies
most probably in the different symmetry group of the horizontal
surface. (For tilted surfaces, the particle-hole symmetry is broken.)

Let us now turn to the discussion of the phase transition from
weak-pinning to strong-pinning regime.

From Fig. \ref{fig:5} we can see, that $\alpha=0$ is stable
fixed point because 
$s_0<1$ for all $\rho$.
On the other hand, we found that the fixed point $\alpha=1$ is unstable for
$\rho=1/2$ 
($s_1>1$), 
but stable for $\rho<1/2$ ($s_1<1$).

The existence of a phase transition for some $\alpha=\alpha_c$ would
be signalled by the presence of a third (unstable) fixed point. We may
formulate it 
by saying 
that $R(\alpha)=\alpha_2(\alpha)-\alpha$ has a root at $0<\alpha_c<1$
with positive derivative, $R(\alpha_c)=0$, $s_c=R^\prime (\alpha_c)>0$. 
The critical value $\alpha_c$ separates the weak-pinning (low
$\alpha$) and strong pinning (high $\alpha$) regimes, corresponding to
different scaling of the dip. 
For $\alpha<\alpha_c$ we have $\gamma<1$, for $\alpha>\alpha_c$ we
have $\gamma=1$.

For $\rho=1/2$ we did not find the third fixed point. Instead,
the unstable fixed point at $\alpha=1$ 
plays its role and  we conclude that there is no phase transition for
$\alpha<1$ and we are in the weak pinning regime for all
$\alpha<1$. 

For $\rho<1/2$ we find $\alpha_c<1$, but very close to 1. The 
values of $\alpha_c$ are 0.9371, 0.9605, 0.97535 for $\rho=1/3, 1/4, 1/5$,
respectively. The dependence of $R(\alpha)$ on $\alpha-\alpha_c$ near
the critical point is shown in the inset in Fig. \ref{fig:5}.

Finally, we compare some of our results with numerical simulations.
We performed direct Monte Carlo simulations of growth in the
inhomogeneous single-step model for horizontal surface, {\it i.e.},
density $\rho = 1/2$ in asymmetric exclusion model.
We measured the dip (\ref{eq:dip}) as the function of 
the system size $L$ and time $t$ for several strengths of 
the defect $\alpha$. When
we start from the flat configuration,
the dip increases with time and after some time $t_{\rm sat}$
(growing with the system size) the stationary regime is reached.
In this regime the dip fluctuates around the characteristic
value $d_{\rm sat}(L)\sim L^{\gamma}$.
We concentrate on the weak defect regime, i.e., small $\alpha$.
In this case fluctuations are very strong and many independent runs are
needed to obtain the value of the saturated dip.
Results presented in 
Fig. \ref{fig:7}
\slaninaefi{
\begin{figure}[hb]
   \centering
   \vspace*{80mm}
   \includegraphics{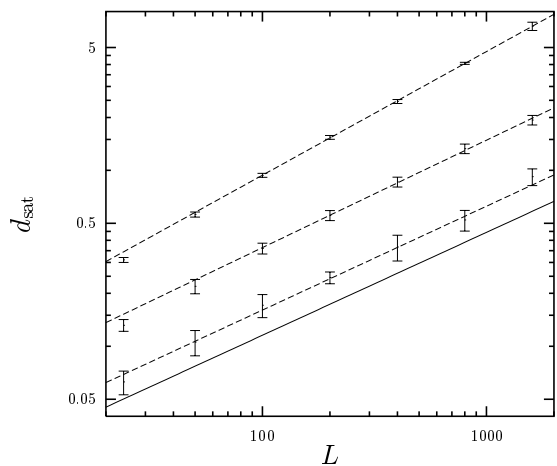}
   \caption{
         Saturated dip,  $d_{\rm sat}$, as a function of the system
         size, $L$, calculated in the inhomogeneous single-step model
         for three strengths of the defect $\alpha=0.05$, 
         $\alpha=0.1$
         and  $\alpha=0.2$ (from bottom to top). The dashed lines are
         fits to last five points, the slope of the full line
         corresponds to the value of $\gamma$ obtained by the
         analytical solution, 
         eq. (\ref{eq:gamma}). }
   \label{fig:7}
\end{figure}
}
are averages over $10^4$ to $5\times10^4$
independent runs.

We observed that the slope, {\it i.e. } exponent $\gamma$
is slightly decreasing with decreasing $\alpha$, but it is definitely larger
than $1/2$. 
We calculated exponent $\gamma$ by fitting the last five points
in each curve.
Values obtained for $\alpha=0.05$, $\alpha=0.1$ and  $\alpha=0.2$
are $0.59\pm 0.03 $, $0.61\pm 0.01$ and $0.703\pm0.004$ respectively.
The value obtained for the lowest value $\alpha=0.05$ agrees well with the 
analytical prediction (\ref{eq:gamma}), so we conclude that the
$\alpha\to 0$ limit is well described by our analytical calculations.

\section{Conclusions and discussion}
We solved exactly in a special case the problem of growing interface
in presence of a point defect. In the language of asymmetric exclusion
model, we obtained the solution for cyclic boundary conditions in the
two-particle sector. The solution consists in two steps. At first, using
the Bethe-Ansatz eigenfunctions   
as a basis set, we found that  the perturbation becomes localized in
the configuration 
space. Then we solved the Dyson equation exactly by inversion of
small 
matrices. We computed the stationary profile of the surface and
determined the dip at the defect. The two-particle result was then
employed in a 
renormalization-group consideration, based on a lattice constant
doubling transformation. We obtained the dependence of the
effective defect strength $\alpha_2$ on the
original, bare defect strength $\alpha$. 
Fixed points of this mapping were identified.

Our main result is the existence of the weak pinning phase and the
value of the corresponding exponent.
Let us discuss first the situation for a horizontal surface ($\rho =
1/2$). 
We have found just two fixed points: $\alpha=0$ which is stable, and
corresponds to weak pinning of the surface to the defect, $d\sim
L^\gamma$ with $0<\gamma<1$, and $\alpha = 1$ which is
unstable. Hence, from our analytical calculations we obtained for
$\alpha<1$ only the weak pinning phase. The strong pinning regime is
limited to a single point $\alpha=1$. We calculated the exponent
$\gamma$ in the weak pinning phase to be $\gamma = 0.58496$.

We verified the existence of weak pinning by a direct Monte Carlo
simulation of the inhomogeneous single-step growth model. For weak
defect, $\alpha$ close to 0, we obtained $\gamma = 0.59 \pm 0.03$ in
good agreement with the analytical prediction. 

The Ref. \cite{wo_ta_90} reports linear dependence of the dip on $L$,
suggesting that $\gamma=1$. In fact, the authors of \cite{wo_ta_90}
state that $d\sim sL$, where $s$ is the slope of the interface at the
defect. They use the parameter $s$ as a measure of the strength of the
defect. We understand this choice as being dictated by the use of
a continuous model (KPZ equation). However, we use different
parametrization, namely taking the 
strength of the defect $\alpha$ as a control parameter. For $\alpha$
fixed, the value of $s$ may depend on $L$. A direct comparison of
Ref. \cite{wo_ta_90} with our results would require measurement of the
quantity $s$, which was not performed in our simulations.

In the case of tilted surfaces, we found that the fixed point $\alpha
= 1$ becomes stable, and in addition there is a third, unstable fixed
point $0<\alpha_c<1$ which corresponds to the phase transition between
strong and weak pinning. However, the value of $\alpha_c$ in our
calculations 
is very close to 1.

The absence of phase transition at $\alpha<1$ for $\rho=1/2$ is in
contradiction with 
numerical simulations on the inhomogeneous polynuclear growth model
\cite{ka_mu_92} as well as  
with analytical results \cite{he_schu_94,ja_le_94}. The most
probable reason lies 
perhaps in the approximation used in calculation of the map
$\alpha_2(\alpha)$: instead of working with macroscopic sample with
generic length $L$ and particle number $n$ and compare it by the 
doubling procedure to 
the sample with length $L/2$ with $n/2$ particles, very small sample,
$L=4$ was used. With such small sample, the influence of boundary
conditions may be very strong, deforming the renormalization map
$\alpha_2(\alpha)$.   
Nevertheless, the agreement of the analytical results with simulations
in the weak pinning phase shows that for weak defects the method we used
works well.

In order to improve the result, it would be
necessary to solve the dynamics for a larger number of particles.
However, the use of the Bethe-Ansatz
eigenfunctions for number of particles larger than two does not lead to
localization of the perturbation and Dyson equation cannot
be solved 
exactly by explicit matrix inversion.  

It would be of interest to investigate the existence of weak pinning
in different inhomogeneous models, in particular one would like to
know if the value of the exponent $\gamma$ for the strength of the
defect going to zero is universal, and eventually if a weak-pinning to
depinning transition exists.

\vspace{5mm}
{\large\bf Acknowledgments}
The work was partially supported by the grant No. A 1010513 of the GA
AV\v{C}R. We wish to thank J. 
Ma\v{s}ek for many useful discussions.


\end{document}